\documentclass[12pt]{article}
\usepackage{epsf}
\usepackage{amsmath}
\usepackage{graphics}
\usepackage{cite}

\renewcommand{\thefootnote}{\#\arabic{footnote}}
\begin{document}

\newcommand{\gtrsim}{ \mathop{}_{\textstyle \sim}^{\textstyle >} }
\newcommand{\lesssim}{ \mathop{}_{\textstyle \sim}^{\textstyle <} }

\newcommand{\rem}[1]{{\bf #1}}

\renewcommand{\thefootnote}{\fnsymbol{footnote}}
\setcounter{footnote}{0}
\begin{titlepage}

\def\thefootnote{\fnsymbol{footnote}}

\begin{center}
\hfill hep-th/0703162\\
\hfill March 2007\\
\vskip .5in
\bigskip
\bigskip
{\Large \bf Entropy of Contracting Universe in Cyclic Cosmology}

\vskip .45in

{\bf Lauris Baum and Paul H. Frampton}

{\em Department of Physics and Astronomy,}

{\em University of North Carolina at Chapel Hill, NC 27599-3255, USA}

\end{center}

\vskip .4in
\begin{abstract}
Following up a recent proposal \cite{BF} for a cyclic model based
on phantom dark energy, we examine the
content of the contracting universe (cu) and its entropy $S_{cu}$. We 
find that beyond dark energy the universe contains on average 
zero or at most a single photon which if present immediately after 
turnaround has infinitesimally energy
which subsequently blue shifts
to produce $e^+e^-$
pairs. These statements are independent of the
equation of state $\omega = p/\rho$ of dark energy
provided $\omega < -1$. 
Thus $S_{cu} = 0$ and 
if observations confirm $\omega < -1$ 
the entropy problem is solved. We discuss the absence of 
a theoretical lower bound on $\phi = |\omega + 1|$, then describe
an anthropic fine tuning argument that renders unlikely 
extremely small $\phi$. The present bound $\phi \lesssim 0.1$ already implies a 
time until turnaround of 
$(t_T - t_0) \gtrsim 100$ Gy.    
\end{abstract}
\end{titlepage}

\renewcommand{\thepage}{\arabic{page}}
\setcounter{page}{1}
\renewcommand{\thefootnote}{\#\arabic{footnote}}

\newpage

In a recent paper \cite{BF} we suggested a cyclic model
which solves the entropy problem \cite{Tolman}. Technical
calculations are certainly desirable to demonstrate the
consistency of our model but we know of no fatal flaw.
The most important new ingredient 
is the idea that the contracting
universe has essentially zero entropy and comes back empty
of matter. Here we take the model seriously 
\footnote{It has
been said 
the problem is not that theoretical physicists take
their own model too seriously but that they
do not take it seriously enough\cite{Weinberg}.} 
and examine more critically some general features including its
possibilty of being tested.

The contracting universe of the cyclic model contains dark energy 
with zero entropy and possibly, as was only generally stated in
\cite{BF}, a small amount of
radiation which could possess entropy. The 
deflation at turnaround 
reduces entropy from a gigantic value O($> 10^{88}$) to an extremely
low value there cited as O($10^1$). An unrealistic value for the dark energy
equation of state $\omega = p/\rho = -4/3$ was
employed simply for algebraic simplicity as it makes
$\rho_{\Lambda} \propto a$, and no attempt was made at
a realistic description of our universe. 
In the present article,
we shall study the entropy of the contracting
universe in this speculative scenario more quantitatively and now will
use arbitrary $\omega = -1 -\phi$ with $\phi > 0$ so that
$\rho_{\Lambda} \propto a^{3 \phi}$. 

The quantity $\phi$ is the most important parameter
for observational discrimination between this cyclic model
and a cosmological constant
\footnote{and from the Steinhardt-Turok cyclic model
\cite{ST1,ST2,ST3}.}.
The next test of $\phi \neq 0$
will likely come from the
Planck Surveyor satellite\cite{Planck}. One wonders,
therefore, how different from zero $\phi$ is? There
is no lower bound on $\phi$ to make the
model work except that it must be non zero.
We already know $\phi \lesssim 0.1$ from the
WMAP3 data \cite{WMAP3}. If $\phi$ is truly infinitesimal, the
test must await improved technology.
To restore optimism we shall, at the end of the Letter,
describe an anthropic
fine tuning argument that
shows that extremely small $\phi$ is unlikely.

In \cite{BF}, it was emphasized that the universe comes back
empty of matter including black holes. The presence
of matter during contraction causes apparently
insuperable problems because accelerated structure
formation will precipitate a premature bounce. Black
holes, if present, will expand and proliferate
with the same consequence.
But the presence of radiation must also be carefully
studied because
although at turnaround the photon energy is
infinitesimal ($E_{\gamma} \lesssim 10^{-200} eV$),
the blue shifting during contraction
leads before the bounce
to production of $e^+e^-$ pairs, undesirable
because generically they will create problems
with continued contraction. As we shall show herein
there are fortunately no photons in the contracting
phase of the cycle, only the truly innocuous dark energy. 

\bigskip
\bigskip
\bigskip

The cyclic model contains one free parameter, the
common density $\rho_C$ at which the universe both turns around
and bounces. Since the bounce is independent of $\omega$ we
begin with it and take as bounce temperatures $T_B = 10^p$ GeV
with, to be above the weak and below the Planck scales, $3 \leq p \leq 17$.
Using the derivation in \cite{BF} this gives
$\rho_C = \eta \rho_{H_2O}$ where $\eta = 10^{(19+4p)}$
and $\rho_{H_20} = 1 g/cm^3$ is the density of water, an easily imaginable
unit somewhere between the unimaginably small
present mean cosmic density and the unimaginably large critical density
$\rho_C$ at turnaround and bounce.

\bigskip

Going now to the turnaround at time $t=t_T$ the scale factor $a(t_T)$
is given by (since $a(t_0)=1$ and putting $\rho_0 = 10^{-29} \rho_{H_2O}$)
\begin{equation}
a(t_T)^{3\phi} = 10^{29} \eta = 10^{48+4p}
\label{tT}
\end{equation}
The present radiation temperature is $(T_{\gamma})_0 = 2 \times 10^{-4}$ eV,
and so from Eq.(\ref{tT}) the radiaition temperature at turnaround
is
\begin{equation}
(T_{\gamma})_T = 2 \times 10^{-4} 
\left( 10^{(48+4p)} \right)^{- 1/3\phi} {\rm eV}
\label{TgammaT}
\end{equation}
which is infinitesimal: putting $\phi=0.1$, Eq.(\ref{TgammaT})
gives $10^{-200}$ eV for p=3 and $10^{-390}$ eV for p=17;
with $\phi = 0.01$, the photon energy is $10^{-2000}$ eV for p=3
and $10^{-3900}$ eV for p=17. In all cases, the photon
wavelength is an astronomical number of orders
of magnitude longer than the present Hubble length.

\bigskip

To evaluate the contraction entropy
we need to estimate many such photons are 
in one causal patch at turnaround. As shown in \cite{BF}
the deflationary factor multiplying entropy 
at turnaround must be much less than the inverse
of the inflationary increase ($\gtrsim 10^{84}$)
of the early universe. 
We take the huge number of causal patches
to be $10^{90} \alpha$ where $\alpha \gg 1$ is a 
parameter to allow an arbitrarily larger number,
and $\alpha=1$ will give an overestimate of contraction entropy.

\bigskip
\bigskip

At turnaround the scale factor is
\begin{equation}
a(t_T) = \left( 10^{(48+4p)} \right)^{\frac{1}{3 \phi}}
\label{aT}
\end{equation}
so taking the present volume as $10^{84} cm^3$ and the present
radiation density as $\rho_r(t_0) = 10^{-33} g/cm^3 = 1 eV/cm^3$
gives for the radiation energy in one causal patch
\begin{equation}
(E_r)_{patch} = \frac{1}{(100 \alpha)^3} \left( 10^{(48+4p)}
\right)^{- \frac{1}{3 \phi}} ~~~ {\rm eV}
\label{Epatch}
\end{equation}

\bigskip

Comparison with Eq.(\ref{TgammaT}) then gives for the number
of photons per causal patch

\begin{equation}
n_{\gamma} = \frac{1}{200 \alpha^3} \ll 1
\label{ngamma}
\end{equation}
which is small even for the unrealistic case $\alpha = 1$
and essentially zero for $\alpha \gg 1$. Thus, the entropy
of the contracting universe (cu) vanishes $S_{cu} = 0$
for any value of equation of state of the dark energy
$\omega = p/\rho = -1-\phi$ since Eq.(\ref{ngamma})
has no $\phi$ dependence.

\newpage

\bigskip
\bigskip

\noindent {\it Anthropic fine tuning argument about $\phi$}

\bigskip
\bigskip

\noindent The time until turnaround is given,
{\it e.g.} \cite{FT}, by
\begin{equation}
(t_T-t_0) \simeq  \frac{t_0}{\phi}
\label{tremaining}
\end{equation}
so if we take, for simplicity, the origin of life to
have occurred at $t_0$ after the most recent bounce
we see from Eq. (\ref{tremaining}) that given small
$\phi \ll 1$ then $\phi$ measures
the fraction of the expansion phase taken to originate
life. An anthropic argument is: it is unreasonable for
the fraction $\phi$, assuming it is non zero,
to be extremely close to zero.

The special case $\phi=0$ is the standard cosmological model
with a cosmological constant where there is no turnaround
and the future lifetime is infinite so the origin of life
necessarily takes place after a vanishing fraction of the
expansion lifetime. Although such an infinite expansion
seems to us unaesthetic \cite{BF}, not all colleagues
share our concern.

As soon as one commits to $\phi \neq 0$, however, the
anthropic type argument emerges and it is unlikely
that $\phi <<< 1$. For example, if $\phi = 10^{-3}$
the length of the expansion phase is $10^4$ Gy whereas
life orinated after only about 10 Gy which is only
0.1\% of the expansion time. If life plays a central role in
our universe, as in our understanding is the spirit
of the anthropic principle, such a tiny value
of $\phi$ is strongly disfavored; one expects
at least $\phi \gtrsim 0.01$ so the fraction before
the origin of life is $ \gtrsim 1.0\%$ of the total
expansion time.

This encouraging argument
makes it more optimistic that the next generation
of observations such as the Planck Surveyor \cite{Planck}
will succeed in detecting a $\phi \neq 0$.

\bigskip
\bigskip
\bigskip

\begin{center}

{\bf Acknowledgements}

\end{center}

\bigskip
This work was supported in part by the
U.S. Department of Energy under Grant No. DE-FG02-06ER41418.

\end{document}